\documentclass{ws-p9-75x6-50}

\begin{document}

\title{Pulsar Timing Observations and Tests of General 
Relativity in Double-Neutron-Star Binaries}

\author{I. H. Stairs}

\address{University of Manchester, Jodrell Bank Observatory,
Macclesfield, Cheshire SK11 9DL U.K.\\E-mail: istairs@nrao.edu}

%%%%%%%%%%%%%%%%%%%%%%%%%%%%%%%%%%%%%%%%%%%%%%%%%%%%%%%%%%%%%%
% You may repeat \author \address as often as necessary      %
%%%%%%%%%%%%%%%%%%%%%%%%%%%%%%%%%%%%%%%%%%%%%%%%%%%%%%%%%%%%%%

\maketitle

\abstracts{We describe the techniques used in pulsar timing observations,
and show how these observations may be applied to tests of
strong-field general relativity for double-neutron-star binary
systems.  We describe the tests of GR resulting from the PSRs B1913+16
and B1534+12 systems.  For the latter pulsar, 5 "Post-Keplerian"
timing parameters are measurable, including the orbital period
derivative and the two Shapiro delay parameters.  }

Pulsar timing is a two-stage process.  First, the pulse profile
obtained in an individual observation is cross-correlated with a
``standard profile'' for the pulsar in question, built up from many
hours' observation; the offset between the two profiles is added to
the starting timestamp of the observation to produce a high-precision
time of arrival (TOA).  The collection of TOAs is then fit to a
phase-coherent timing model, which includes the pulsar position,
period, period derivative and dispersion measure.  For binary pulsars,
five Keplerian orbital parameters (orbital period, projected orbital
semi-major axis, orbital eccentricity, longitude and epoch of
periastron) must also be fit.  This model fitting is usually
accomplished with the program {\sc tempo} (see
http://pulsar.princeton.edu/tempo).  The two stellar masses are then
the only unknown system parameters.

In the case of certain binary pulsar systems, notably the
double-neutron-star binaries, ``post-Keplerian'' (PK) parameters can
also be fit.  These include the rate of periastron advance,
$\dot\omega$, the time-dilation and gravitational-redshift parameter,
$\gamma$, the rate of orbital period decay, $\dot P_b$, and the
Shapiro-delay parameters, $r$ and $s$.  In practice, these parameters
are fit in a theory-independent manner.\cite{dd86} Within a given
theory of gravity, each PK parameter can be written as a function of
the two stellar masses.  Thus if two PK parameters are measured, the
system is completely determined within a particular theory of gravity,
while if three or more are measured, the system is overdetermined and
can be used to test the theory itself.\cite{tw89,twdw92}

The best-known example of the use of a double-neutron-star system to
test gravitational theories is PSR~B1913+16, discovered at Arecibo in
1974.\cite{ht75a}  For this system, the $\dot\omega$, $\gamma$ and
$\dot P_b$ parameters are measured, and agree with the predictions of
general relativity to better than 1\%.\cite{tw89,dt91}  Through the
measurement of $\dot P_B$, this system provided the first proof of the
emission of gravitational radiation as predicted by the theory of
general relativity.\cite{tay94b}

A similar double-neutron-star system, discovered at Arecibo in 1990,
is PSR~B1534+12.\cite{wol91a} This pulsar permits the measurement of
$\dot\omega$, $\gamma$ and $\dot P_b$, as for PSR B1913+16, but also,
because of the favourable inclination of the binary orbit, the two
Shapiro delay parameters $r$ and $s$.  Furthermore, PSR~B1534+12 has a
stronger, sharper pulse than PSR~B1913+16, thereby permitting more
precise TOAs to be calculated; it may eventually surpass PSR~B1913+16
in the precision of its timing solution.  At the present time,
$\dot\omega$, $\gamma$ and $s$ are measured extremely precisely, and
also agree with the predictions of general relativity to better than
1\%.\cite{sac+98}  It should be noted that the agreement between
these PK parameters represents a test of the purely quasi-static
regime of general relativity, complementing the mixed quasi-static and
radiative test obtained from PSR~B1913+16.  The range of Shapiro
delay, $r$, is also in agreement with general relativity despite a
large measurement uncertainty; this should improve with further timing
observations.

The orbital period derivative of PSR~B1534+12 merits special
discussion.  The observed value of $\dot P_b$ cannot be compared
directly to the value predicted by general relativity.  The relative
acceleration of the Solar System Barycenter and the center of mass of
the pulsar system induces a bias in the observed value.  In the case
of PSR~B1534+12, the bias is dominated by a proper motion term
dependent on the distance of the pulsar from the earth.\cite{shk70} At
present, the only, and rough, estimate of the distance to the pulsar
comes from its dispersion measure and a model of the free electron
content in the galaxy,\cite{tc93} which yield an estimate of
0.7$\pm$0.2\,kpc.  When the bias is calculated using this value for
the distance, the resultant measured intrinsic $\dot P_b$ is
$(-0.167\pm0.018)\times 10^{-12}$, only 87\% of the value predicted by
GR, $-0.192\times 10^{-12}$.\cite{sac+98} Under the assumption that GR
is the correct theory of gravity, the test can in fact be inverted and
an improved distance to the pulsar calculated;\cite{bb96} by this
method, the distance to PSR~B1534+12 is found to be
1.1$\pm$0.2\,kpc.\cite{sac+98} An independent measurement of the
pulsar distance, through a timing or interferometric parallax, will
automatically lead to a greatly strengthened test of general
relativity from this pulsar system.

\section*{Acknowledgments}
I thank my colleagues in the PSR B1534+12 observations: Zaven
Arzoumanian, Fernando Camilo, Andrew Lyne, David Nice, Joe Taylor,
Stephen Thorsett and Alex Wolszczan.  I also thank Michael Kramer for
reading the manuscript.

\end{document}